\documentclass[11pt]{article}
 \usepackage{graphicx}

\renewcommand{\vec}[1]{{\bf #1}}

\def\beq{\begin{eqnarray}}
\def\eeq{\end{eqnarray}}

\def\al{\alpha}
\def\be{\beta}

\def\de{\delta}
\def\vp{\varepsilon}

\def\na{\nabla}
\def\pa{\partial}

\def\La{\Lambda}

\def\Om{\Omega}

\begin{document}

\hfill {Preprint number: DF-UFJF-01-06-08}

\hfill ArXiV:  0806.1969 [astro-ph]

\vskip 6mm

%%%%    \hfill hep-th/
%%%%    \vskip 5mm
%%%%%%%%%%%%%%%%%%%%%%%%%%%%%%%%%%%%%%%%%%%%%%%%%%%%%%%%%%%%
%
\begin{center}
{\Large\sc DM particles: how warm they can be?}
\vskip 5mm

{\small \bf Julio C. Fabris $^{(a)}$}
%% \footnote{E-mail address: fabris@cce.ufes.br}
\ , \qquad
{\small \bf Ilya L. Shapiro $^{(b)}$} \footnote{Also at
Tomsk State Pedagogical University, Russia.}
%% E-mail address: shapiro@fisica.ufjf.br}
\ , \qquad
{\small \bf Fl\'avia Sobreira $^{(c,b)}$}
%% \footnote{E-mail address: flavia$\_$sobreira@yahoo.com.br}
\vskip 4mm

%%%%%%%%%%%%%%%%%%%%%%%%%%%%%%%%%%%%%%%%%%%%%%%%%%%%%%%%%%%%%%%%%
%%%%    Julio C. Fabris, Ilya L. Shapiro, Flavia Sobreira    %%%%
%%%%%%%%%%%%%%%%%%%%%%%%%%%%%%%%%%%%%%%%%%%%%%%%%%%%%%%%%%%%%%%%%

(a) \ Departamento de F{\'\i}sica -- CCE, Universidade Federal
do Esp{\'\i}rito Santo 
\\
Vit\'oria, CEP: 29060-900, ES, Brazil
\vskip 2mm

(b) \ Departamento de F\'{\i}sica -- ICE,
Universidade Federal de Juiz de Fora
\\
Juiz de Fora, CEP: 36036-330, MG,  Brazil
\vskip 2mm

(c) \ Instituto de Física Te\'orica, Universidade Estadual Paulista
\\ 
Rua Pamplona 145, Bela Vista, S\~ao Paulo, CEP: 01405-000, SP, Brazil 

\end{center}

\begin{quotation}

%%%%%%%%%%%%%%%%%%%%%%%%%%%%%%%%%%%%%%%%%%%%%%%%%%%%%%%%%%%%%%
%%%%%%%%%%%%%%%%%%%%%%%%%%%%%%%%%%%%%%%%%%%%%%%%%%%%%%%%%%%%%%
\vskip 2mm

%%%%%%%%%%%%%%%%%%%%%%%%%%%%%%%%%%%%%%%%%%%%%%%%%%%%%%%%%%%%
\noindent
{\large\bf Abstract.} \
One of important questions concerning particles which compose 
the Dark Matter (DM) is their average speed. We consider the 
model of relativistic weakly interacting massive particles 
and try to impose an upper bound on their actual and past 
warmness through the analysis of density perturbations and
comparison with the LSS data. It is assumed that the DM can 
be described by the recently invented model of reduced 
relativistic gas (RRG). The equation of state of the RRG 
model is closely reproducing the one of the Maxwell 
distribution, while being much simpler. This advantage 
of the RRG model makes our analysis very efficient. As a 
result we arrive at the rigid and model-independent bound 
for the DM warmness without using the standard (much more 
sophisticated) approach based on the Einstein-Boltzmann 
system of equations. 
\vskip 2mm

%%%%%%%%%%%%%%%%%%%%%%%%%%%%%%%%%%%%%%%%%%%%%%%%%%%%%%%%%%%%%
{\bf Pacs:} \ {98.80.-k, 98.80.Cq, 98.80.Bp}
%%%%%%%%%%%%%%%%%%%%%%%%%%%%%%%%%%%%%%%%%%%%%%%%%%%%%%%%%%%%%%%%%%%
%% 98.80.-k  Cosmology                                            %
%% 98.80.Bp  Origin and formation of the Universe                 %
%% 98.80.Cq  Particle-theory and field-theory models of the       %
%%     early Universe (including cosmic pancakes, cosmic strings, %
%%     chaotic phenomena, inflationary universe, etc.)            %
%%%%%%%%%%%%%%%%%%%%%%%%%%%%%%%%%%%%%%%%%%%%%%%%%%%%%%%%%%%%%%%%%%%

\vskip 2mm

%%%%%%%%%%%%%%%%%%%%%%%%%%%%%%%%%%%%%%%%%%%%%%%%%%%%%%%%%%%%%%%%%%%
{\bf Keywords:} \ Warm Dark Matter, Ideal Relativistic Gas, 
Cosmic Perturbations.
\vskip 2mm
\end{quotation}

%%%%%%%%%%%%%%%%%%%%%%%%%%%%%%%%%%%%%%%%%%%%%%%%%%%%%%%%%%%%%
%%% \vskip 4mm                                              %
%%%                                                         %
%%%%%%%%%%%%%%%%%%%%%%%%%%%%%%%%%%%%%%%%%%%%%%%%%%%%%%%%%%%%%
%%%%%%%%%%%%%%%%%%%%%%%%%%%%%%%%%%%%%%%%%%%%%%%%%%%%%%%%%%%%%
\section{\large\bf  Introduction}                           %
%%%                                                         %
%%%%%%%%%%%%%%%%%%%%%%%%%%%%%%%%%%%%%%%%%%%%%%%%%%%%%%%%%%%%%

%%%%% $\,$ \quad 
The Sun is shining bright in Brazil. However, independently
on geography, cosmologists say the Universe is dominated by 
a darkness. Namely, the energy balance of the present-day 
Universe shows that the relative energy densities of the Dark 
Energy and Dark Matter (DM) are close to $\Om_\La^0=0.7$ 
and $\Om_{DM}^0=0.25$, respectively, while the visible (more 
precise, baryonic) matter is represented by a modest less 
than 5\% of the total energy density \cite{hannestad}.

Despite the existing variety of the models for Dark Energy,
the Cosmological Constant (CC) $\La$ is the most natural 
candidate. The presence of a $\La$-term is dictated by the
requirement of consistency of quantum field theory in curved
space. At the same time the enormous fine-tuning which is
necessary for adjusting the value of $\La$ to the astronomical
observations creates a longstanding CC problem (see, e.g.,
\cite{weinberg89,nova} for discussion). However, in this paper 
we will concentrate on the second dark component, which is 
equally mysterious. The main candidate to be DM is the gas of
weakly interactive massive particles (WIMPs) which could be 
part of a multiplet composition of some extension of the 
Standard Model of elementary particle physics. For example, 
those can be superpartners of observable particles in MSSM or 
in some supergavity model. One can find an extensive discussion 
of the DM issue in the books \cite{KT,Dod,Mukh,CoLu} or in the 
recent reviews \cite{DM1,DM2,susyDM}.

In simplest terms one can describe the DM problem as 
follows. The astronomical observations show that the stars 
and interstellar gas clouds in the spiral 
galaxies have the rotation curves different from the ones 
produced by gravitational field of the visible matter. 
The typical spiral galaxy has an almost flat structure, 
while the gravitational field is apparently produced by some 
almost spherical distribution of mass, total amount of it 
should be a few times greater than the one of the visible part. 
The hidden mass presumably forms a halo and is called DM. The 
main question is from what the DM is made. Obviously, the 
constituents of the DM should have properties distinct from 
the ones of the baryonic matter, for otherwise the two kinds 
of matter would be distributed in the same way. Furthermore, 
in the cosmological setting, DM is necessary for the cosmic 
structure formation.

One can distinguish the three main kinds of DM. The first 
one is cold DM, e.g., formed by WIMPs. Another one is  
hot DM, which can be represented, e.g., by massive neutrinos 
(with a mass of some $eV$). Hot dark matter leads to the 
so-called up-bottom scenario, where structures of clusters 
of galaxies are formed first, while cold dark matter implies 
the bottom-up scenarios, forming first small objects of 
scales smaller than a galaxy. Even if the cold dark matter 
scenario seems more favored, each scenario has its own 
problems, with suppression (hot DM) or excess (cold DM) 
of power at small scales \cite{ostriker1}. The intermediate 
scenario of warm DM has been invoked to solve this problem
\cite{szalay,ostriker2,ostriker3,dodelson,muller}. Warm DM 
may be composed, e.g., by relatively heavy sterile neutrino 
with the $keV$-scale mass (see, e.g., 
\cite{first,dodelson,DM7,Shap}) or
by some other particles such as light gravitinos \cite{DM8}.
The structure formation in these models has been explored 
using fluid description \cite{muller} (see also \cite{dodelson}
and references therein) and also $N$-body simulations methods 
\cite{DM6,DM2}. 

%%%%%%%%%%%%%%%%%%%%%%%%%%%%%%%%%%%%%%%%%%%%%%%%%%%%%%%%%
The concepts of hot and warm DM imply that the DM constituents 
are relativistic in the early Universe and then cool down when 
the Universe expands. Qualitatively similar evolution takes 
place also for the baryonic matter, however the last emits 
radiation and therefore cools down faster. 
Due to the growing amount of the available experimental 
data, the requirements for the cosmological models are 
becoming stronger and in particular it would be desirable 
to have more precise description of the expansion of the 
Universe. In both cases of baryonic and DM one needs to have 
a model which could take into account,  in a natural way, 
the continuous evolution of the equation of state of the 
matter content. In other words, this should be a model 
with the radiation-like behavior in the early Universe, 
becoming more like a dust or like a DM in the later epoch. 

In the present paper we start the detailed analysis of the 
new, relatively simple model of reduced relativistic gas (RRG). 
This model is reproducing the equation of state of the 
Maxwell distribution with high precision \cite{FlaFlu}. 
The short technical introduction to the RRG model is 
postponed for the next section, but some general discussion 
is in order here. 
The RRG model enables one to consider the matter content 
which is hot in the early Universe and continuously cools 
down when the Universe expands. Let us note that the same 
purpose can be achieved by taking, e.g., relativistic 
distributions for the ideal gas case, such as the Maxwell, 
Fermi-Dirac or Bose-Einstein ones. However, in all these 
cases the relation between the energy density and the 
pressure have rather complicated form. In particular, for 
the most simple Maxwell distribution this relation has the 
form of the ratio of the two modified Bessel functions 
\cite{Juttner} (see also, e.g., \cite{Kremer}). Needless 
to say that such expressions are 
difficult to deal with in the cosmological framework. It 
is important to remember that the equation of state is the 
unique relevant element of the matter content, as far as we 
are interested only in the behavior of the conformal factor 
of the metric  (zero-order cosmology). In this respect
the RRG model \cite{FlaFlu} provides a real advantage. 
Due to its simplicity, one can easily integrate the 
Friedmann equation and develop the zero-order cosmological 
model in the economic and analytic form. From the other 
side, as we have already mentioned above, the RRG model 
provides the equation of state which is very close to the 
one corresponding to the Maxwell distribution. Therefore 
the properties of the RRG-based cosmological model can be 
safely attributed to the model of the Universe filled by 
an ideal gas of massive relativistic particles. 

In the framework of the RRG model one arrives at the 
single-fluid cosmology which interpolates, 
in a natural way, between the radiation-like and 
dust-like regimes. Therefore, at the level of zero-order 
cosmology the RRG model is proved to be a useful tool. 
Even more important, one can generalize the RRG model by
implementing the possibility of the energy exchange between 
RRG gas and other fluids by introducing viscosity (see, 
e.g., \cite{FWC} and references therein). In this way we 
may expect to obtain the most precise zero-order cosmology 
which can be hopefully given by an analytical solution 
and be very useful. In 
future, we may have a chance to see the impact of different 
interaction types of the DM particles, e.g., on the cosmic 
perturbations spectrum. 

However, before the RRG model can be considered as a tool 
for creating the precise cosmological model, it has to be 
submitted to another important test. The present-day 
cosmological model does not deserve confidence if it 
produces good results only at the zero-order level. So, 
the next step in the development of the RRG model is to 
check whether it can produce acceptable results for the 
cosmic perturbations. If the RRG model can pass this test, 
it is worthwhile to use it as a building block for 
constructing the realistic cosmological models in a 
way described above. For this reason, in the present 
paper we address the issue of density perturbations 
spectrum in the simple RRG model \cite{FlaFlu}.  
%%%%%%%%%%%%%%%%%%%%%%%%%%%%%%%%%%%%%%%%%%%%%%%%%%%%%%%%%%%%
The output of the analysis of the cosmic perturbations in the 
RRG model can be compared to existing detailed description 
of the perturbations in the warm and hot DM models (see, 
e.g., \cite{szalay,ma} and also \cite{DM1,DM2,Dod,Mukh}
for the review). Indeed, our purpose is not to compete 
with the standard results, which are based on the numerical 
solution of the much more detailed description in the 
framework of the complicated Einstein-Boltzmann system. 
Instead of this, we want to see whether the results 
derived within the RRG model are compatible with the 
standard ones and, in this way, to try our model. 

Let us remember that the RRG model is reproducing the ideal 
relativistic gas, which is isotropic and, therefore, can 
not provide full information on the motion of the 
DM particles. In the framework of such simplified 
approach we have a restricted choice of relevant 
physical observables, the most obvious is an average 
speed of the DM particles. Hence, our immediate purpose 
here is to establish an upper bound for the velocities of 
the DM constituents, both in the present and earlier epochs 
of the Universe. The upper bound for the DM velocities comes 
from the fit with the LSS data \cite{cole}. In this way, we 
can test to which extent the dark matter can be hot or at 
least warm. It is well known that the standard way to 
impose the bound on the warmness of the DM particles 
is through the analysis of cosmic perturbations in the 
Einstein-Boltzmann coupled system \cite{ma}. As we shall 
see in what follows, the use of the relatively simple RRG 
model enables one to achieve similar restrictions in a much 
more economic way. Thus, using this approach, we circumvent 
the technical difficulties related to the analysis of the 
Einstein-Boltzmann system without losing the essential 
features. 
%%%%%%%%%%%%%%%%%%%%%%%%%%%%%%%%%%%%%%%%%%%%%%%%%%%%%%%%%

The paper is organized as follows. In the next section 
we present a very brief introduction to the RRG model. 
The reader can find further details in \cite{FlaFlu}. 
In section 3 the equations for density perturbations and 
their numerical analysis are considered and in section 4 
we present some discussions and draw our conclusions.

%%%%%%%%%%%%%%%%%%%%%%%%%%%%%%%%%%%%%%%%%%%%%%%%%%%%%%%%%
%%%%%%%%%%%%%%%%%%%%%%%%%%%%%%%%%%%%%%%%%%%%%%%%%%%%%%%%%
% Section 2                                             %
\section{\large\bf Reduced model for relativistic gas}  %
%                                                       %
%%%%%%%%%%%%%%%%%%%%%%%%%%%%%%%%%%%%%%%%%%%%%%%%%%%%%%%%%

The equation of state for the ideal relativistic gas of 
identical massive particles has been derived in 
\cite{Juttner}. This equation involves a ratio of two 
modified Bessel functions and is rather difficult to 
apply for the cosmological purposes. One can simplify 
things considerably if assuming that, instead of the 
Maxwell law, all particles have equal kinetic energies.
An elementary consideration leads to the following
relation between pressure $P$ and energy density
$\rho=n\vp$:
\beq
P = \frac{\rho}{3}\,\cdot\,
\left[\,1-\Big(\frac{m c^2}{\vp}\Big)^2\,\right]\,,
\label{state 1}
\eeq
where $\vp=mc^2/\sqrt{1-\be^2}$, \ $n$ is a number of
particles per unit of volume and $\be=v/c$.
One can introduce the new notation for the density of
the rest energy 
\beq
\rho_d=nmc^2\,,
\label{state r}
\eeq 
where $n$ is the number of particles for a unit of $3d$ 
volume. This density depends on the scale factor in the 
usual way
$$
\rho_d = \frac{a_0^3}{a^3}\,\rho_d^0\,,
$$
where $\rho_d^0$ is the rest energy density at present, 
when $a=a_0$. Using this quantity, one can rewrite 
Eq. (\ref{state 1}) in the form
\beq
P = \frac{\rho}{3}\,\cdot\,
\left[\,1\,-\,\frac{\rho_d^2}{\rho^2}\,\right] \,.
\label{state 2}
\eeq
Both forms (\ref{state 1}) and (\ref{state 2}) will 
be useful for us in what follows.
An important observation is that the expression
(\ref{state 2}) reproduce the Maxwell-based equation
of state with very good precision. According to the 
plot obtained in \cite{FlaFlu}, the maximal difference 
between the equations of 
state $\,\rho=\rho(P)\,$ in the two cases is just 
2.5\% of the absolute value of $\rho$ and, moreover,
this discrepancy goes to zero pretty fast in the UV, 
when $\rho \gg \rho_d$.

Let us emphasize that the difference between the 
equation of state which follows from relativistic 
Maxwell distribution and the equation of state in 
our simplified model is so small that it can be seen
as negligible, when we use this equation of state, e.g., 
in the Friedmann equation. Therefore, the cosmological 
model which we are going to develop on this background, 
will be based on the following two assumptions: 
\ 1) that the massive particles (e.g. 
the ones of DM or baryonic matter) go from one equilibrium 
state to another in a sufficiently smooth way, such that 
the fluid composed by these particles can be described by 
the equation of state instead of the Boltzmann equation
(if compared to the standard approach \cite{ma}). 
\ 2) That the interaction between these particles is
negligible. Indeed, the main advantage of our model is 
that it enables one to introduce interactions between 
the particles in a very elegant way. We shall treat this 
issue in a separate paper and now concentrate on the 
ideal gas case. 

Since the RRG model is really close 
to the Maxwell distribution, in what follows we shall 
refer to the velocity of the particles in the RRG as to 
``average speed''. This term will help us to keep in 
mind that the results of our calculations provide the 
reliable information not only about the proper RRG 
model, but also about the Maxwell-distributed 
relativistic gas. 

Using the conservation law 
\beq 
- \,\frac{d\rho}{\rho + P}\,=\,\frac{3da}{a}
\label{scale 1} 
\eeq 
and the equation of state (\ref{state 2}), 
one can easily arrive at the RRG density scaling rule 
\beq 
\rho\,=\,\rho(z)
\,=\,\frac{\rho_c^0\,\Om_M^0}{\sqrt{1 + b^2}}
\,\big(1+z\big)^3\,\sqrt{1 + b^2\,\big(1+z\big)^2}\,, 
\label{scale 3}
\eeq 
where $\Om_M^0=\Om_{DM}^0+\Om_{BM}^0$ is a total relative
present-day matter energy density, $\rho_c^0$ is the 
present day critical density and $z=-1+a_0/a$ is the 
red-shift parameter. The dimensionless parameter $b$ 
shows whether the 
velocity of the RRG particles is large or small or, in 
other words, whether the matter is ``cold'', or ``warm '', 
or ``hot''. In order to better understand the physical 
sense of this parameter, let us express it in two 
different (albeit equivalent) forms as follows: 
\beq 
b = \frac{\rho^0_d}{\rho^0} = \frac{\be}{\sqrt{1 - \be^2}}\,. 
\label{b} 
\eeq
Indeed, $b\approx 0$ means that the particles are 
nonrelativistic and that the RRG is nothing but the dust.  
Furthermore, for small velocities one can just set 
$b=\be$. The main purpose of this paper is to establish 
the upper bound for the parameter $b$ from the analysis 
of cosmic perturbations.

It is easy to see that the expressions (\ref{state 1}),
(\ref{state 2}) and (\ref{b}) interpolate between the dust 
(the limit $b=0$) and radiation ($b\to \infty$) extreme 
cases. It is important to note that the expression 
(\ref{scale 3}) does not represent a simple sum of the 
pressureless and radiation components. Instead this is 
a formula for the ideal relativistic gas of massive 
particles which undergoes an adiabatic expansion. At 
high red shift $z\to \infty$ the gas is compressed and 
its temperature is high. In this case the expression 
above looks as ultrarelativistic one. When $\,z\to -1$, 
the gas becomes almost pressureless and the above 
expression is close to the one for the dust. Of course, 
we are interested in the finite time intervals and, for 
this reason, can not separate the RRG equation of state 
and scale dependence (\ref{scale 3}) to the radiation 
and dust parts. Due to this feature the expression above 
looks as a useful tool for various problems of cosmology. 
In particular, here we will use the RRG model as a new 
tool for testing the warmness of the DM today and 
in the early Universe. 

Let us consider the cosmological solution in the RRG 
model \cite{FlaFlu}. The Friedmann-Lema\^{\i}tre 
equation has the form 
\beq
H^2(z)=\frac{8\pi G}{3}\,
\big[\rho(z)+\rho_\La\big]+H_0^2\Om_k^0(1+z)^2\,, 
\eeq 
where
$\rho(z)$ is given by (\ref{scale 3}) and 
$\rho_\La=\La/8\pi G$.
This equation can be presented in the explicit form 
\beq
H^2\,=\,H_0^2\,\Big[\Om^0_k\,(1+z)^2\,+\,
\frac{\Om^0_M}{\sqrt{1+b^2}}\, (1+z)^3\,\sqrt{1\,+\,
b^2(1+z)^2}\,+\,\Om_\La^0\Big]\,. 
\label{hubble} 
\eeq 
Let us introduce the following two useful notations:
\beq 
g(z) \,=\,\frac{(1+z)\,H}{3[H^2- \Om_k^0 H_0^2 (1+z)^2]} 
\,,\qquad
f_1(z) \,=\,\frac{\rho(z)}{\rho_t(z)}
\,=\, \frac{(1+z)(H^2)^\prime 
- 2\Om_k^0 H_0^2\,(1+z)^2} {[H^2-\Om_k^0 H_0^2(1+z)^2]\,(4-r)}\,,
\label{function} 
\eeq
where the prime means derivative $d/dz$. In the last formulas 
we denoted the ratio of the square of the rest energy density 
(\ref{state r}) and the energy density $\rho$ as  
$$
r = r(z)=\frac{\rho^2_d(z)}{\rho^2(z)}\,,
$$ 
and also applied the usual sum rule 
$\Om^0_M+\Om_\La^0+\Om_k^0=1$.

The cosmological model based on RRG with the presence of 
the cosmological constant admits an analytic solution 
\cite{FlaFlu}. This solution, as one should expect, does 
interpolate between the ones for the radiation 
and the dust cases. In the very early Universe, when the 
temperature is high, the evolution of the Universe is 
close to the one in the radiation - dominated case. 
At the other end of the energy scale, in the late Universe, 
the solution becomes close to the one for the pressureless 
matter case. \ Such interpolation between the two regimes 
is qualitatively similar to the more conventional case 
where the matter content is composed by a sum of the dust 
and radiation. In the conventional case, also, radiation 
dominates at high $z$ and dust 
dominates at low $z$. However, in the RRG case we observe 
this property in a cosmological model with a single fluid. 
This property makes RRG model a useful tool for modeling 
the behaviour of a DM particles in the expanding Universe. 
The analytic zero-order solution can be, in principle,
extended for the combination of the relativistic gas of 
massive particles and radiation, or for the combination 
of several distinct RRG-like fluids (for example, this 
can be done by using the method of \cite{Analit-IFT}).

%%%%%%%%%%%%%%%%%%%%%%%%%%%%%%%%%%%%%%%%%%%%%%%%%%%%%%%%%
%%%%%%%%%%%%%%%%%%%%%%%%%%%%%%%%%%%%%%%%%%%%%%%%%%%%%%%%%
% Section 3                                             %
\section{\large\bf Perturbations spectrum}              %
%%%%%%%%%%%%%%%%%%%%%%%%%%%%%%%%%%%%%%%%%%%%%%%%%%%%%%%%%

Consider the cosmic perturbations in the RRG model 
described above. We shall follow the scheme elaborated 
for the exploration of another model which is motivated 
by quantum corrections \cite{CCwave} and consider 
simultaneous perturbations of metric, energy density 
and the 4-velocity in the co-moving coordinates
\beq 
U^\al \rightarrow U^\al + \de
U^\al \,, \qquad \rho \rightarrow \rho\,\left(1 + \de\right)\,, 
\qquad
g_{\mu\nu} \rightarrow  g_{\mu\nu} + h_{\mu\nu} \label{plus} 
\eeq
In the synchronous coordinates we have $h_{0\mu}=0$. The 
perturbation of the pressure should be derived from the eq. 
(\ref{state 1}), such that 
$$
\de P= \frac{\de \rho\,(1-r)}{3}\,.
$$ 
In this way we arrive at the
following $00$-component of the Einstein equation
%%%%%%%%%%%%%%%%%%%%%%%%%%%%%%%%%%%%%%%%%%%%%%%%%%%%%%%%%%%%
%% \beq
%% \dot{\hat h}\,+\,2H\,{\hat h}
%% \,=\,8\pi G\,\big[\,2\,-\,r(z)\,\big]\,\de\,\rho\,,
%% \label{mice}
%% \eeq
%%%%%%%%%%%%%%%%%%%%%%%%%%%%%%%%%%%%%%%%%%%%%%%%%%%%%%%%%%%%
\beq
h^\prime\,-\,\frac{2h}{(1\,+\,z)}
\,=\,-\frac{f_1\,(2-r)}{g} \,\de\,,
\label{first}
\eeq
where \ ${\hat h}=\pa_t\left(h_{kk}/a^2\right)$. 
Other equations follow from the variation of the conservation 
law \ $\de\left(\na_\mu T^\mu_\nu\right)=0$ and have 
the form
\beq
\de^\prime\,-\,\frac{1}{(1+z)}
\,\Big[4-r-\frac{(1+z)\rho^\prime}{\rho}\Big]\,
\de\,\,+\,\,\frac{4-r}{3H(1+z)}\,
\Big(\,\frac{\hat h}{2}\,-\frac{v}{f_1}\,\Big)\,=\,0\,,
\label{pink}
\eeq
and
\beq
v^\prime\,+\,\Big(\frac{\rho^{'}}{\rho}
-\frac{r^\prime}{4-r}-\frac{5}{1+z}
-\frac{f_1^\prime}{f_1}\Big)\,v\,+\,\frac{k^2(1+z)f_1}{H}
\,\,\frac{1-r}{4-r}\,\,\de\,=\,0\,,
\label{out}
\eeq
where \ $v=f_1\left(\na_k \de U^k\right)$ and we used 
Fourier expansions for $\de(z)$ and $v(z)$
\beq
\de({\vec x},z)=\int \frac{d^3k}{(2\pi)^3}\,\,
\de({\vec k},z)\,e^{i{\vec k}\cdot{\vec x}}\,, 
\quad
v({\vec x},z)=\int \frac{d^3k}{(2\pi)^3}\,\,
v({\vec k},z)\,e^{i{\vec k}\cdot{\vec x}}\,, 
\quad
k=\left|{\vec k}\right|. 
\nonumber %% \label{Furier} 
\eeq

In order to explore the equations (\ref{first}), (\ref{pink}) 
and (\ref{out}) one has to choose the initial conditions, 
related to the choice of the transfer function. We have 
performed the numerical analysis using two kinds of these 
functions. The more complicated one was introduced in 
\cite{Bardeen} and was explained in details, e.g., in 
\cite{Martin,CCwave}. The second option is 
a more simple transfer function from the book \cite{CoLu}.
Both transfer functions assume a scale invariant primordial
spectrum, and determine the spectrum today considering the
Universe with the cosmological constant and filled by DM.
Using the
transfer functions we can fix the initial conditions at a 
redshift after the recombination epoch. It is remarkable 
that the results for the power spectrum obtained through 
these transfer functions actually coincide. The most relevant 
plots which show the comparison with the 2dFGRS observational 
data \cite{cole} are presented in Fig. 1. 
%%%%%%%%%%%%%%%%%%%%%%%%%%%%%%%%%%%%%%%%%%%%%%%%%%%%%%%%%%%%%%%%

%%%%%%%%%%%%%%%%%%%%%%%%%%%%%%%%%%%%%%%%%%%%%%%%%%%%%%%
%%%%%%%%%%%%%%%%%%%%%%%%%%%%%%%%%%%%%%%%%%%%%%%%%%%%%%%
\begin{figure}
\includegraphics[width=0.40\textwidth]{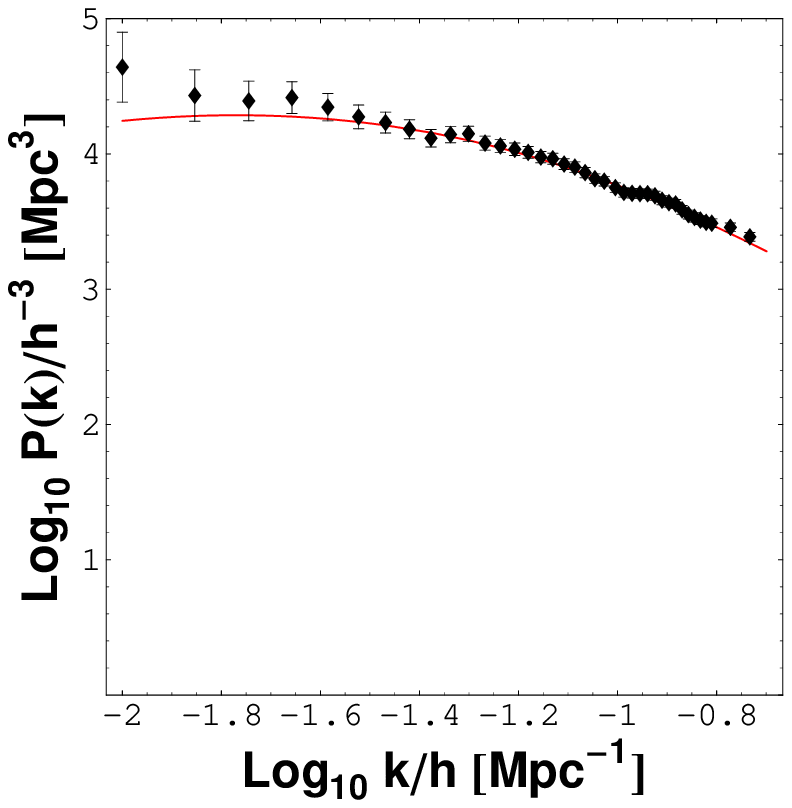}
\qquad\quad
\includegraphics[width=0.40\textwidth]{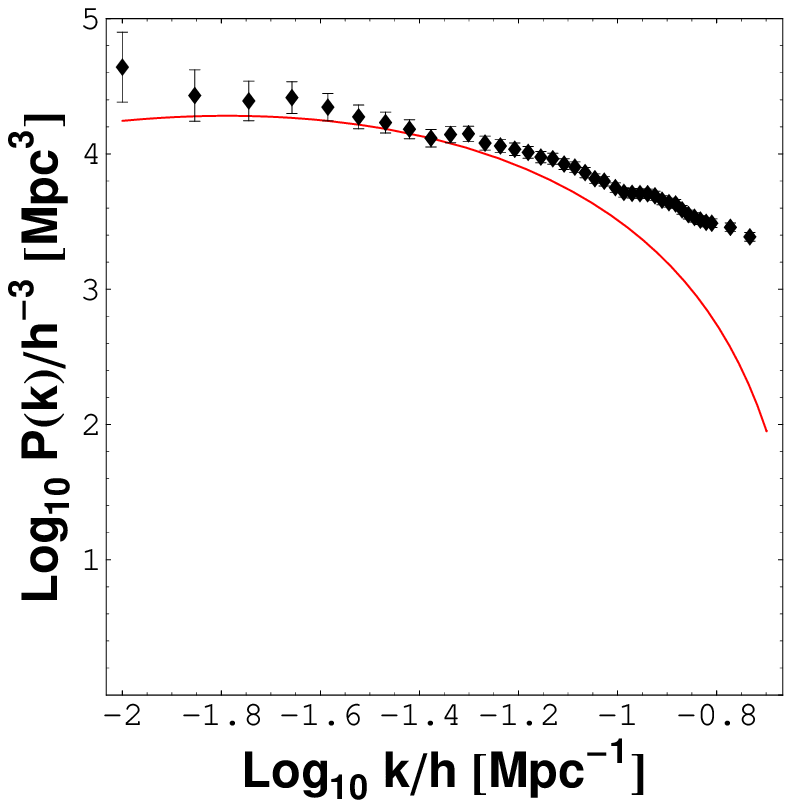}
\\ \\
\includegraphics[width=0.40\textwidth]{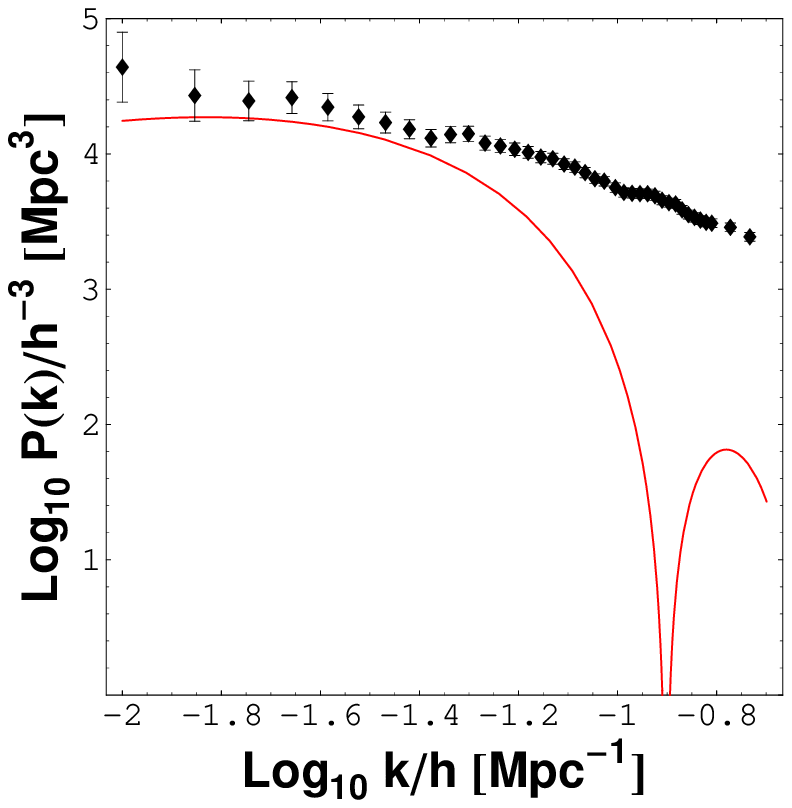}%
\qquad\quad
\includegraphics[width=0.40\textwidth]{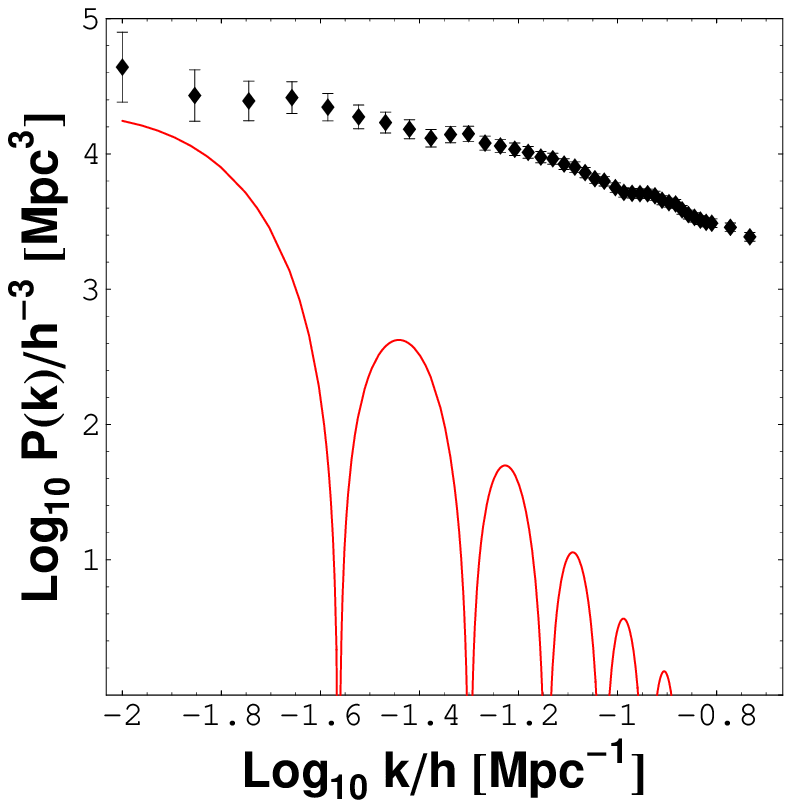}
%% \vskip -14.5cm
%% \hspace*{-7cm}a)\hspace*{8cm}b)
%% \vskip 6.5cm
%% \hspace*{-7cm}c)\hspace*{8cm}d)
%% \vskip 6cm
\caption{Power spectrum for the RRG-$\Lambda$ model, for
fixed $\Omega_{B}^{0}=0.04$, $\Omega_{DM}^{0}=0.21$ and
$\Omega_{\Lambda}^{0}=0.75$ (flat Universe), with
the values \ $b=10^{-5}$,  \ $b=10^{-4}$,
\ $b=2\times 10^{-4}$ \ and  \ $b=10^{-3}$.
The theoretic plots are presented together with the
LSS data from the 2dfFGRS \cite{cole}. The ordinate axis
represents the log of $P(k)=|\delta_m(k)|^2$ at $z=0$.
In the abscissa we have the log of the wave
number $k$ given in $h\,$Mpc$^{-1}$ units.} 
%% \label{fig:pp_tt}}
\end{figure} 
%%%%%%%%%%%%%%%%%%%%%%%%%%%%%%%%%%%
%%%%%%%%%%%%%%%%%%%%%%%%%%%%%%%%%%%

The relevant quantity to be compared with the observational 
data is the power spectrum parameter defined by
\begin{equation}
{\cal P}_k = \delta_k^2 \quad ,
\end{equation}
where $\delta_k$ is the component of the Fourier transform 
of the density contrast $\de(t)$, which is computed by 
integrating the equations for the cosmic perturbations 
(\ref{first}), (\ref{pink}) and (\ref{out}) for a given 
value of $k$ and with a given initial conditions (as 
indicated above). 

In the present case, since the upper bound for the possible 
values of $\,b\,$ has great physical relevance, it proves 
useful to establish this bound with more certainty. For this 
end we performed calculation for a set of different values 
of $b$ and then applied the statistical method to compare 
the result to the power spectrum data of the 2dFGRS 
observational program. 
The quality of the fit between the theoretical estimate and 
the observational data can be characterized by the quantity
\begin{equation}
\chi^2 = \sum_i\biggr(\frac{{\cal P}_{k_i}^o 
-{\cal P}_{k_i}^t}{\sigma_i}\biggl)^2 \,,
\end{equation}
where ${\cal P}_{k_i}^o$ is the observational data for the 
power spectrum for a given wavenumber $k_i$, 
\ ${\cal P}_{k_i}^t$ 
is 
the corresponding theoretical result obtained by the 
numerical integration of the equations for the 
perturbations (\ref{first}), (\ref{pink}), (\ref{out}) 
and $\sigma_i$ are the observational error bars related 
to the measurement. As smaller is the parameter $\chi^2$, 
better is the fit. Of course, since our theoretical model 
depends on some input parameters such as \ $b$, 
\ $\Omega_M^0$ and $\Omega_\La^0$, the value of $\chi^2$ 
depends also on these parameters. 
At a first step, in order to obtain estimations for the 
relevant parameters, we reduce the three-dimensional 
probability distribution to the one-dimensional one 
by choosing the values of the present day 
cosmological parameters
$$
\Om_M^{0} = \Om_{B}^{0} + \Om_{DM}^{0} 
= 0.04  + 0.21 = 0.25 
\qquad \mbox{and} \qquad
\Om_{\Lambda}^{0}=0.75\,,
$$ 
which correspond to the flat space section of the space-time 
manifold. 

Using the quantity $\chi^2$, the probability distribution 
is given by
\beq
\label{prob}
P = A\,e^{- \chi^2/2} \quad ,
\eeq
where $A$ is a normalization constant. The final result for
the one-dimensional probability distribution function (PDF) 
for the parameter $b$ is displayed in Fig. 2. Using this 
plot we can see that this probability distribution goes 
sharply to zero for the values above $b = 5\times10^{-5}$. 
We can conclude that there is an upper bound for the
``warmness'' parameter $b$, that means 
\ $b\leq 3\times 10^{-5} \, \sim 4\times 10^{-5}$.
%%%%%%%%%%%%%%%%%%%%%%%%%%%%%%%%%%%%%%%%%%%%%%%%%%%%%%%%%%

%%%%%%%%%%%%%%%%%%%%%%%%%%%%%%%%%%%%%%%%%%%%%%%%%%%%%%%%%%
\vskip 5mm
\begin{figure}
\includegraphics[width=0.5\textwidth]{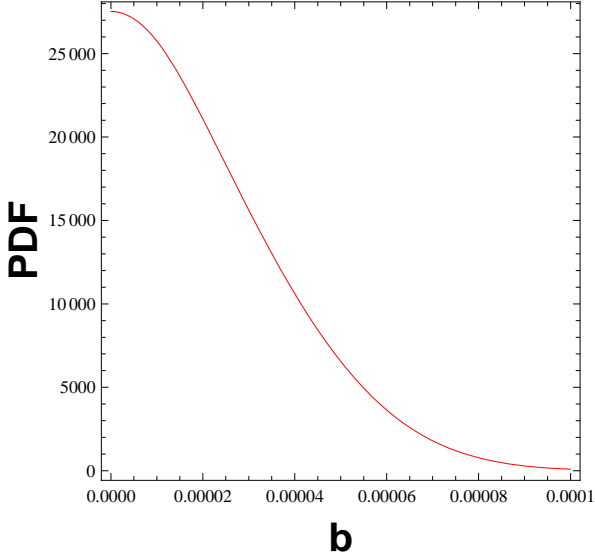}
\caption{Probability distribution for the parameter $b$. 
The probability becomes essentially zero for 
$b \geq 5\times10^{-5}$.}
\end{figure}
%%%%%%%%%%%%%%%%%%%%%%%%%%%%%%%%%%%%%%%%%%%%%%%%%%%%%%%%%%
%%%%%%%%%%%%%%%%%%%%%%%%%%%%%%%%%%%%%%%%%%%%%%%%%%%%%%%%%%

%%%%%%%%%%%%%%%%%%%%%%%%%%%%%%%%%%%%%%%%%%%%%%%%%%%%%%%%%%
%%%%%%%%%%%%%%%%%%%%%%%%%%%%%%%%%%%%%%%%%%%%%%%%%%%%%%%%%%

Before starting to discuss physical interpetations of 
our results, let us present some comments concerning the 
validity and restrictions of the bound on the parameter 
$\,b\,$ obtained above. 
%%%%%%%%%%%%%%%%%%%%%%%%%%%%%%%%%%%%%%%%%%%%%%%%%%%%%%%%%%
The power spectrum, which is defined in the space of the 
$k's$, is the Fourier transformation of the two-point 
correlation function, which is defined in the real space, 
$\vec x$. Hence, the power spectrum implies a Fourier 
decomposition. The different Fourier modes are independent 
only when the linear approximation is valid; non-linearity 
leads to the mixing of the different Fourier modes, which 
do not remain independent anymore. The non-independence of
Fourier modes must be taken into account when the statistics
analysis in the $k$ space is performed. This is done through 
the use of the covariance matrix, $C_{ij}$. According to 
this method, the likehood function for a given set of data 
$\Delta_i$ is given by
\beq
{\cal L} \propto 
\exp\biggr\{- \frac{1}{2}\Delta_i C_{ij}^{-1}\Delta_j\biggl\}\,,
\eeq
where $C_{ij}^{-1}$ is the inverse of the covariance matrix. If
the conditions of linearity are satisfied, the covariance matrix
becomes diagonal, and the likehood function reduces to the usual
expression encoded in the $\chi^2$ parameter.
\par
We will use the 2dFGRS data such that 
$0 < k\,h^{-1} < 0.185\,Mpc^{-1}$ \cite{cole}. 
This interval lies outside the non-linear
regime, generally fixed as $k\,h^{-1} \stackrel{>}{\sim}
0.8\,Mpc^{-1}$, corresponding to a scale greater than $8\,Mpc$. 
If the linear approximation is valid, the covariance matrix is
diagonal, and the use of a $\chi^2$ statistic is justified.
However, the limits on the wavelength where the linear 
regime can be safely applied is not very well established. 
In reference \cite{percival}, it has been argued that the 
confidence on the linear approximation for the 2dFGRS data
restricts the scales to $k\,h^{-1} \leq 0.15\,Mpc^{-1}$. On the
other hand, the cosmic variance leads to a restriction such that
$k\,h^{-1} \geq 0.02\,Mpc^{-1}$. If we use data ranging in the
interval $0.02\,Mpc^{-1} \leq k\,h^{-1} \leq 0.15 \leq Mpc^{-1}$
the results does not change substantially, as it will be discussed
latter.

To obtain a better estimate for the velocities of the DM 
particles in halo of galaxies, we must enter deeply into the
non linear regime. All considerations presented in this work are
based on the applicability of the linear analysis and therefore
we assume it to be valid. Using the matter power spectra data, 
we can not avoid in this case the use of the full covariance 
matrix, since the different modes are not independent 
anymore (see for example reference \cite{szapudi} for 
a beautiful analysis of the non-linear effects and the 
consequent use of the covariance matrix in the context of 
the baryonic acoustic oscillations). Hence, our previous 
estimations must be seen as a lower bound, since the process 
of contraction in the formation of the galaxy must increase 
the velocities of the particles.

\begin{center}
\begin{figure}[!t]
\begin{minipage}[t]{0.5\linewidth}
\includegraphics[width=\linewidth]{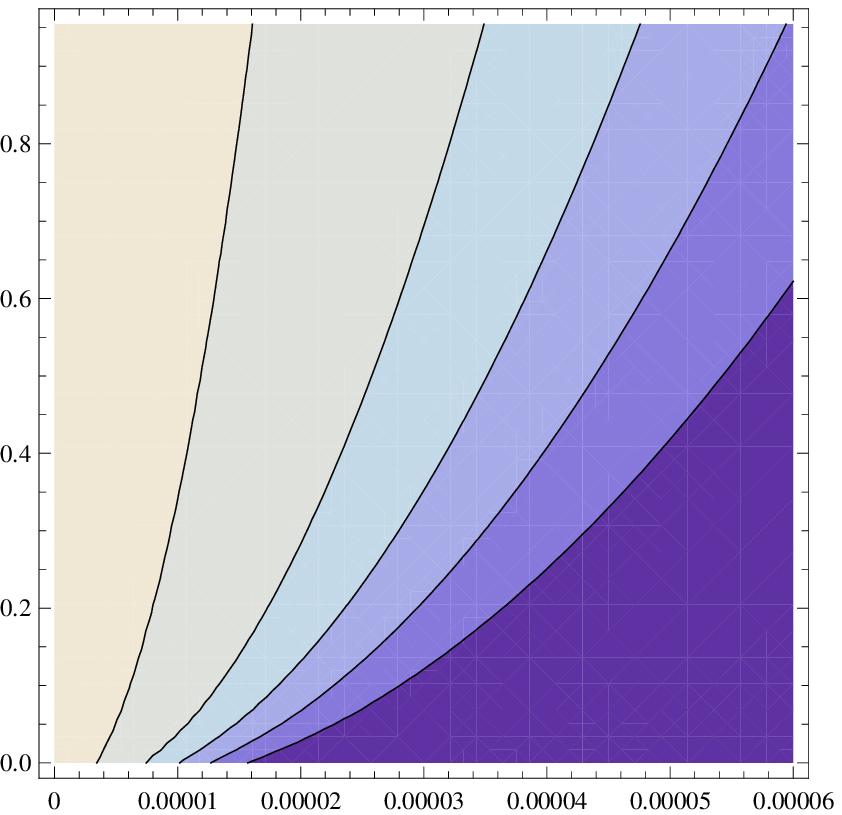}
\end{minipage} \hfill
\begin{minipage}[t]{0.5\linewidth}
\includegraphics[width=\linewidth]{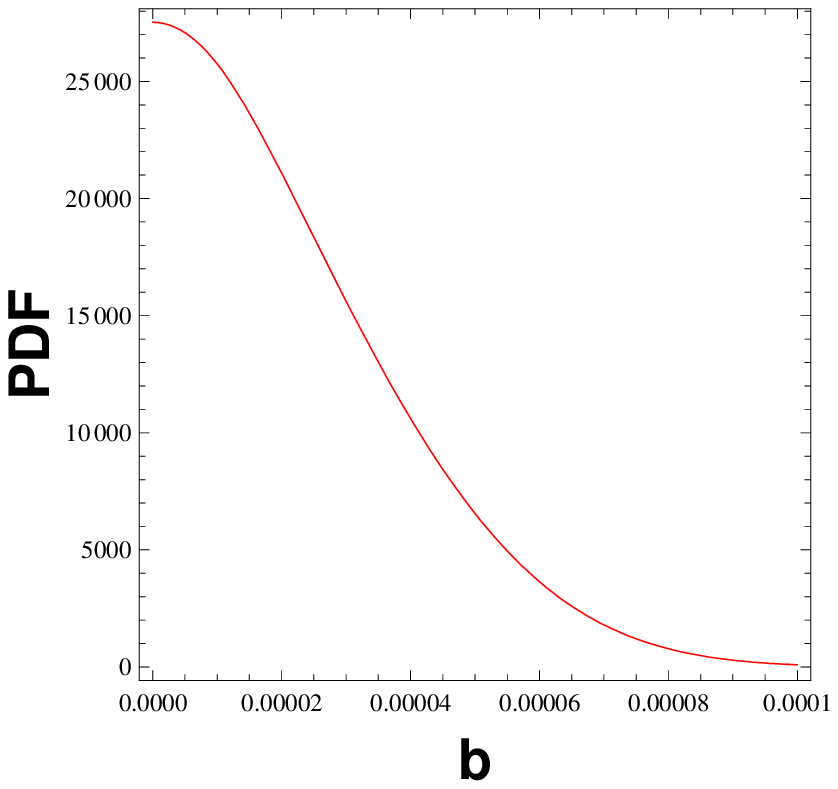}
\end{minipage} \hfill
\caption{{\protect\footnotesize The two-dimensional probability
distribution, for the flat spatial section, when both $b$ and
$\Omega_{\Lambda0}$ are varied (left), considering the modes $0
\leq kh^{-1} \leq 0.185 Mpc^{-1}$. 
Higher probabilities are indicated by the brighter regions. 
The one-dimensional probability
distribution for $b$, after marginalizing on $\Omega_{\Lambda0}$
is shown at the plot on the right.}} \label{2dmbis}
\end{figure}
\end{center}

\begin{center}
\begin{figure}[!t]
\begin{minipage}[t]{0.5\linewidth}
\includegraphics[width=\linewidth]{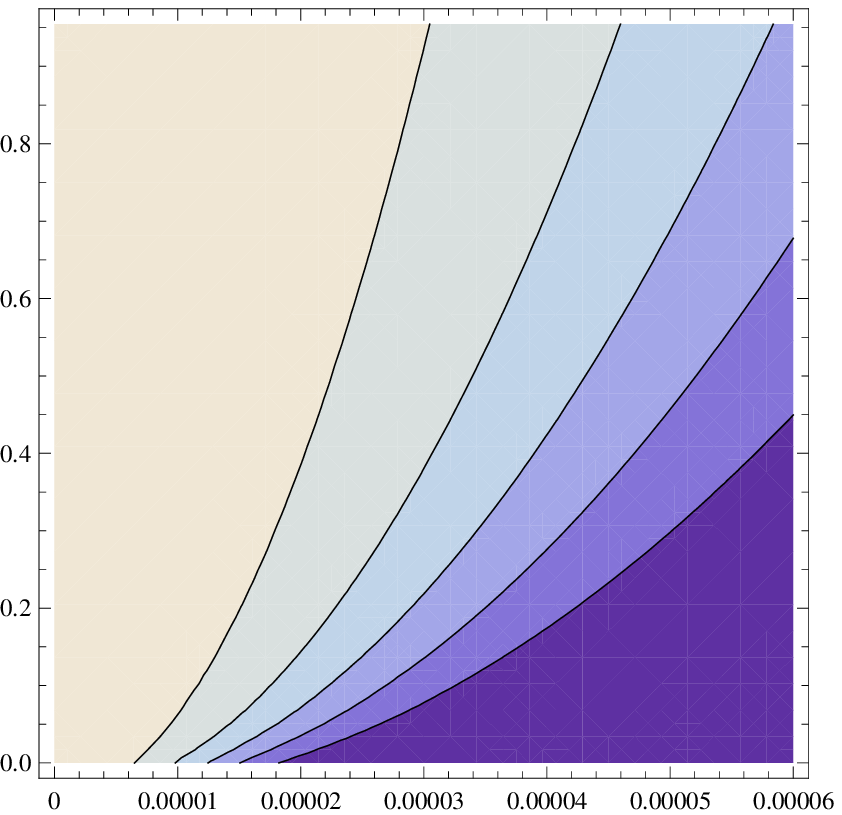}
\end{minipage} \hfill
\begin{minipage}[t]{0.5\linewidth}
\includegraphics[width=\linewidth]{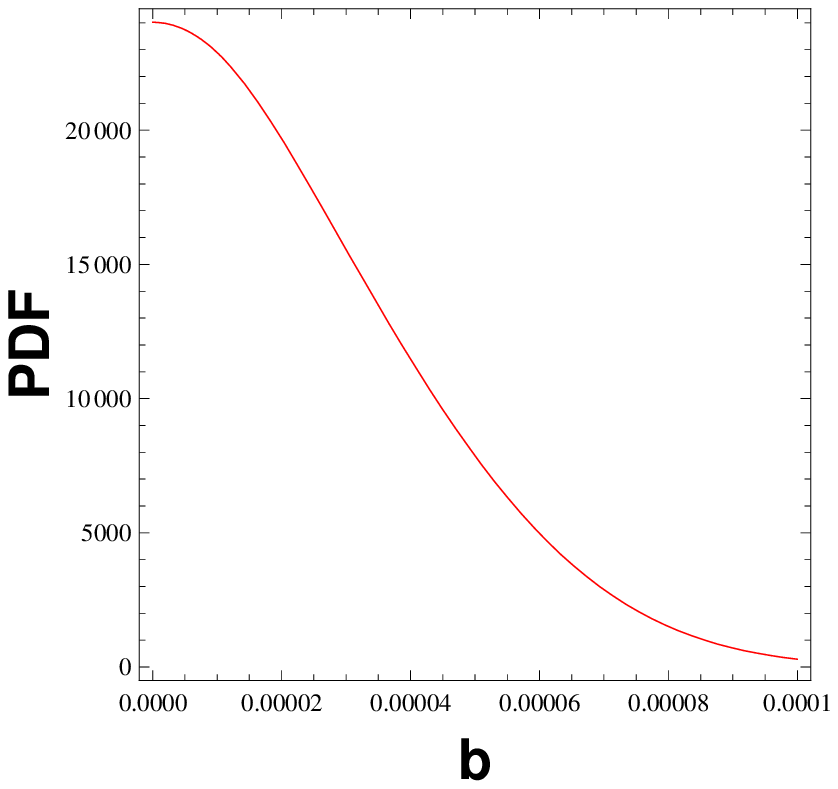}
\end{minipage} \hfill
\caption{{\protect\footnotesize The two-dimensional probability
distribution, for the flat spatial section, when both $b$ and
$\Omega_{\Lambda0}$ are varied (left), but restricting the modes
to $0.2 Mpc^{-1} \leq kh^{-01} \leq 0.150 Mpc^{-1}$. 
Higher probabilities are indicated by the brigther regions. The
one-dimensional probability distribution for $b$, after
marginalizing on $\Omega_{\Lambda0}$ is shown at right.}}
\label{2dm}
\end{figure}
\end{center}

Furthermore, 
in the previous analysis, we have fixed the quantity of dark 
matter and dark energy, based on the 5-years results of Wmap. In
order to verify how this restriction may influence the  obtained 
bounds for the parameter $b$, we now consider a two-dimensional 
parameter space, varying at the same time $b$ and the dark 
matter quantity $\Omega_{dm0}$ (or equivalently the dark 
energy density $\Omega_{\Lambda0}$, since we are restricted 
to a flat spatial section).  The results are shown in figure 
\ref{2dm} using the data $0 \leq kh^{-1} \leq 0.185 Mpc^{-1}$, 
and in figure \ref{2dmbis} using 
$0.2 Mpc^{-1} \leq kh^{-01} \leq 0.150 Mpc^{-1}$. 
According to these plots, the bounds for the value of $b$ 
remain essentially the same when we perform these additional 
variations of other cosmic parameters. Moreover, they are 
also in a good agreement with the case when we used only 
the best fit value for $\Omega_{\Lambda0}$ and 
$\Omega_{dm0}$ given by Wmap.

Now we are in a position to discuss physical significance
of the bounds obtained from cosmic perturbations analysis. 
The restriction for the parameter $b$ which we derived from 
the numerical analysis of cosmic perturbations, can be 
easily translated into the bound for the average velocity 
(or warmness) of the DM constituents. For this purpose we 
have to note that $b$ is necessarily small and therefore
the relation (\ref{b}) converts into $b=\be=v/c$. Then we
arrive at the bound for the average speed of the massive 
relativistic particles of the DM in the present-day Universe
$v \leq v_0$, \ $v_0 \approx 10-12\, km/s$. This bound 
agrees with the standard evaluations (see, e.g.,
\cite{Mukh,KT}) obtained from the numerical simulations 
of the structure formation in the neutrono-dominated 
Universe and also from the model-dependent considerations. 

%%%%%%%%%%%%%%%%%%%%%%%%%%%%%%%%%%%%%%%%%%%%%%%%%%%%%%%%%%%%%
%%%%%%%%%%%%%%%%%%%%%%%%%%%%%%%%%%%%%%%%%%%%%%%%%%%%%%%%%%
%%%%%%%%%%%%%%%%%%%%%%%%%%%%%%%%%%%%%%%%%%%%%%%%%%%%%%%%%%
%%%%%%%%%%%%%%%%%%%%%%%%%%%%%%%%%%%%%%%%%%%%%%%%%%%%%%%%%%

%%%%%%%%%%%%%%%%%%%%%%%%%%%%%%%%%%%%%%%%%%%%%%%%%%%%%%%%%%
%%%%%%%%%%%%%%%%%%%%%%%%%%%%%%%%%%%%%%%%%%%%%%%%%%%%%%%%%%
%%%%%%%%%%%%%%%%%%%%%%%%%%%%%%%%%%%%%%%%%%%%%%%%%%%%%%%%%%

It looks tentative to compare this bound for the speeds 
of the DM particles with some astronomic observable, as
it is usually done in the framework of standard approaches
\cite{ma}. One can, for instance, try to compare this bound 
to the known one for the spiral galaxies, which is 
about\footnote{This approach is close to the one of 
\cite{Drukier}, which is based on the Maxwellian dark 
matter velocity distribution for spherically symmetric 
and isotropic halo.} $\,240\,km/s$. \ Obviously, there 
is no correspondence between the two numbers. However, 
let us note that the galaxy is an object which definitely 
lies out of the linear perturbation regime which we deal 
with here. In general, any comparison of the results 
obtained from the linear approximation to the 
cosmological perturbations, including the one for 
the upper limit of the velocity of DM particles with 
the dynamics of virialized systems in the Universe 
(galaxies, clusters of galaxies, etc.) must be performed 
with great caution, because those virialized objects are 
in the deep non-linear regime, with typical densities 
hundreds of times larger than the critical density. 

In general, the estimations of velocity of dark matter 
particles strongly depend on the nature of the DM candidate 
under consideration and especially on the reference frame 
where the velocity is evaluated. For example, in the paper
\cite{sikivie} were obtained a dispersion of primordial 
velocities (which is not affected by non-linear effects) 
with respect to the Hubble flow. The result is of the 
order of magnitude about $10^{-12}$ in unities of the 
velocity of light for the WIMPS with a typical mass of
a few $GeV$'s and about $10^{-17}$ in the same units 
for the axion, even if 
the typical mass of the axions is some $10^{16}$ orders 
smaller than the typical mass of the WIMPs. This case shows 
how the nature of the dark matter candidate influences 
the estimate for its primordial velocity. In reference 
\cite{abazajian}, the possibility
that the warm dark matter can be described  by sterile 
neutrinos has been analysed. The possible mass range goes 
from $1\,keV$ to tenths of $keV$. Now, using the results 
for the WIMPS quoted above, and the usual non-relavistic 
relation for the ratio of velocities of particles having 
the same energy but different masses, we find an upper 
limit for the velocity of the warm dark matter particle 
of the order of $10^{-9}$ in unities of $c$. This is far 
below our upper limit near the maximum of probability 
distribution. On the other hand, the two estimates 
correspond to distinct reference frames and therefore 
these results do not contradict our analysis.

We could try to use, for instance, the data of weak 
lensing investigations concerning the dark halo of 
galaxies. But, the results obtained in this way still 
concern a very non-linear regime, see for example 
\cite{mellier}. In order to chose the appropriate 
object, we note that among the galaxies, the dwarf 
spheroidal elliptical ones are those with the large 
proportion of dark matter to baryonic matter, with 
a mass/luminosity ratio that can be as large as 
$500$, in solar units, while ordinary galaxies have 
a ratio of the order of some $10-50$ \cite{binney}. 
Even if these objects 
are deep in the non-linear regime it is remarkable that 
the dispersion velocity becomes essentially constant 
far from the center with a typical value of the order 
of $v \sim 10\,km/s$ \cite{walker}, comparable with the 
velocity bounds we have obtained for the dark matter 
particle. 

Since the dwarf spheroidal galaxies represent the extreme 
case of virialized system dominated by dark matter, such 
agreement of the typical velocities with our bound is 
perhaps not meaningless, even if a clear determination 
of the extension of the dark halo would be necessary to 
put this comparison into more solid grounds. For the dwarf 
irregular or spiral galaxies the typical radial speed of 
the stars is evaluated to be about $10-12\, km/s$
\cite{dwarfs} and we meet a nice correspondence with our 
bound. It is amusing that we arrived at this correspondence 
by using a very simple RRG model \cite{FlaFlu} and not the 
complicated approach based on the Einstein-Boltzmann system 
\cite{ma}. 

The last problem which we can easily address within the 
RRG model is the dynamics of the DM average speeds in 
the expanding Universe. In other words, it would be 
interesting to calculate how this speed depends on the 
red-shift parameter $z$, or on the temperature $T_{CMB}$ 
of the cosmic background radiation (CMB). In order to 
address this issue one has to use the relations (\ref{b}) 
and (\ref{scale 3}). The unique role of the parameter 
$b$ is to define the ``warmness'' of the DM in the 
last of these relations, so it is obvious that at the 
higher $z$ we have \ $b(z) = b (1+z)$, where $b$ 
is the modern value. Furthermore, since the upper bound 
for $b$ nowadays is much less than $10^{-4}$, for the 
potentially relevant $z \leq 1000$ we can, according to  
(\ref{b}), safely use the formula \ $v=bc$. In this 
way we arrive at the following relation for the average 
speed of the DM particles
\beq
v(z) \,=\, cb(z) \,=\, c b\,(1+z) \,=\, 
v \times \frac{T_{CMB}}{T_{CMB}^{(0)}}\,,
\label{vbz}
\eeq
where we used $\,T_{CMB} \sim (1+z)\,$ and denoted 
$\,T_{CMB}^{(0)}$,  $\,b\,$  and $\,v\,$ the corresponding 
quantities for $z=0$. So, in the framework of the RRG model 
we note that the average kinetic energy of the DM particles 
is proportional to the square of the CMB temperature, the 
result which is familiar from the conventional (but more
complicated) considerations (see, e.g., \cite{DM2}). 
 
%%%%%%%%%%%%%%%%%%%%%%%%%%%%%%%%%%%%%%%%%%%%%%%%%%%%%%%%%
%%%%%%%%%%%%%%%%%%%%%%%%%%%%%%%%%%%%%%%%%%%%%%%%%%%%%%%%%
% Section 4                                             %
\section{\large\bf Discussions and conclusions}                         %
%                                                       %
%%%%%%%%%%%%%%%%%%%%%%%%%%%%%%%%%%%%%%%%%%%%%%%%%%%%%%%%%

We have considered the structure formation in the model where 
the DM is described by the ideal relativistic gas of identical 
massive particles. Instead of using Maxwell distribution, we 
have employed the RRG model \cite{FlaFlu} which is closely 
reproducing Maxwell distribution and, at the same time, is 
rather simple. As a result we arrive at the strong limit on 
the parameter $b$, which should satisfy the upper bound 
\ $b\leq 3-4 \times 10^{-5}$. According to the relation 
(\ref{b}), this is equivalent to the upper bound on the 
velocities of the DM particles
\ $v \leq v_0 = 3-4\times10^{-5} c = 10-12\,km/s$. 
This restriction is much more severe than, e.g., 
the one discussed earlier in \cite{DMs1,DMs2,DMs3,DMs4} 
on the basis of the nonrelativistic Maxwell distribution 
\cite{DMs2} and is essentially smaller than 
the typical velocities of the stars in the spiral galaxies. 
Also, it is  about two order of magnitude smaller than the 
escape velocities for the spiral galaxies. 

Does our result mean that the actual velocities of DM 
particles can 
not be greater that the mentioned bound $v_0$? An obvious 
answer is no. Let us remember that both DM and baryonic 
matter can acquire an extra kinetic energy {\it after} 
the galaxy starts to form and the linear regime of the 
cosmological perturbations can not be applied. One can 
see the corresponding process as a kind of the usual 
transformation of the potential gravitational energy
into the kinetic one. This process has nothing to do 
with the linear perturbations we have studies here. 
However taking smaller astrophysical objects such as dwarf
galaxies, we arrive at the surprisingly nice correspondence 
between the observed average speeds of the stars in such 
galaxies and our upper bound $v_0$. This correspondence 
shows that the RRG model is, perhaps, the simplest 
way to arrive at the reasonable estimates concerning not 
only the behavior of the conformal factor, but also the 
linear cosmological perturbations. 

%%%%%%%%%%%%%%%%%%%%%%%%%%%%%%%%%%%%%%%%%%%%%%%%%%%%%
The result described above is universal in the sense there 
is no dependence on the origin and properties of the WDM 
constituents. In particular, the restriction on velocities 
does not interfere with the one for the masses of the WDM 
particles, which can be even macroscopic ones. We were 
just treating them at a component of ideal relativistic 
gas and derive restrictions on their velocities. Hence, 
these restrictions apply equally to all known models of 
WDM, see, e.g., refs. \cite{wilczek,DM6,DM7,DM8,DM9}. 
For example, they apply to the models of DM particles 
which do not interact with other matter and with 
themselves, except gravitationally \cite{DM1,grav1,grav2}.
According to our results, even this simplified model can 
produce good predictions for the spectrum of linear 
perturbations, if we assume the ideal gas
equation of state for these weakly interacted particles. 
Another interesting point is that our results show that 
the anisotropy of the DM distribution does not play a 
critical role in the definition of the perturbations 
spectrum and that the equilibrium distributions such 
as the Maxwell one (which is closely reproduced 
by RRG) is, in principle, sufficient to arrive at the 
reasonable estimates for the speed of the DM particles.  
%%%%%%%%%%%%%%%%%%%%%%%%%%%%%%%%%%%%%%%%%%%%%%%%%%%%%%%%%%%%%%%%

Last, but not least. The RRG model may be successfully applied 
for the investigation of more complicated situations, including 
two distinct non-interacting ideal gases \cite{KM} and also may 
be useful in describing the interaction between these gases, 
e.g. through the use of viscosity (see, e.g., \cite{FWC}). 
We expect to explore these issues in the near future. 
In general, our model proved useful in exploring relativistic 
properties of the ideal gas of massive particles, it can be 
applied for solving various problems of gravitational physics.

In order to verify to which extent the model described above 
may correct the excess of power of the $\La$CDM at small 
scales, the non-linear regime may be explored. As a result 
one may hope to achieve a more detailed description of the 
structure formation. At the moment it is unclear whether the 
RRG model can be a useful tool in such case. We hope to 
explore this issue at the consequent stage of our work.

%%%%%%%%%%%%%%%%%%%%%%%%%%%%%%%%%%%%%%%%%%%%%%%%%%%%%%%%%%
\vskip 4mm
%%%%%%%%%%%%%%%%%%%%%%%%%%%%%%%%%%%%%%%%%%%%%%%%%%%%%%%%%%

\noindent
{\large\bf Acknowledgments.}

Authors are grateful to G. Bertone, S. Dodelson, D. Hooper, 
N. Gnedin and Y. Mellier for useful conversations. J.F. and 
I.Sh are thankful for kind hospitality to the Institut 
d'Astrophysique de Paris and to the Fermilab Theoretical 
Astrophysics Group, correspondingly, where part of their 
work has been performed. The work of the authors has been 
supported by CAPES and CNPq (Brazil), FAPEMIG (MG/Brazil) 
and FAPES (ES/Brazil).

%%%%%%%%%%%%%%%%%%%%%%%%%%%%%%%%%%%%%%%%%%%%%%%%%%%%%%%%%%
%%%%%%%%%%%%%%%%%%%%%%%%%%%%%%%%%%%%%%%%%%%%%%%%%%%%%%%%%%
\vskip 4mm
%%%%%%%%%%%%%%%%%%%%%%%%%%%%%%%%%%%%%%%%%%%%%%%%%%%%%%%%%%

%%%%%%%%%%%%%%%%%%%%%%%%%%%%%%%%%%%%%%%%%%%%%%%%%%%%%%%%%%
%                    THE REFERENCES:                     %
%%%%%%%%%%%%%%%%%%%%%%%%%%%%%%%%%%%%%%%%%%%%%%%%%%%%%%%%%%
  %%{99}


\begin{thebibliography}{99}

\bibitem{hannestad} S. Hannestad, Int. J. Mod. Phys. {\bf A21}
(2006) 1938.

\bibitem{weinberg89} S. Weinberg, Rev. Mod. Phys., {\bf 61} (1989) 1.

\bibitem{nova} I.L. Shapiro, J. Sol\`{a},
% {\sl Scaling behavior of the cosmological constant:
% Interface between quantum field theory and cosmology.}
JHEP {\bf 02} (2002) 006.

\bibitem{KT} E.Kolb and M.Turner, {\sl The Very Early Universe}
(Addison-Wesley, New York, 1994).

\bibitem{Dod} S. Dodelson, {\sl Modern Cosmology}
(Academic Press, New York, 2003).

\bibitem{Mukh}
V. Mukhanov, {\sl Physical Foundations of Cosmology}
(Cambridge University Press, 2005).

\bibitem{CoLu}
P.J.E. Peebles, \textit{Physical Cosmology}, (Princeton University
Press, 1993).

\bibitem{DM1} L. Bergstrom,
% Non-Baryonic Dark Matter - Observational Evidence and
% Detection Methods.
Rept. Prog. Phys. 63 (2000) 793, hep-ph/0002126.

\bibitem{DM2}
G. Bertone, D. Hooper and J. Silk,
% Particle Dark Matter: Evidence, Candidates and Constraints
Phys. Rept. 405 (2005) 279, %% -390
hep-ph/0404175.

\bibitem{susyDM}
%% Supersymmetric dark matter.
G. Jungman, M. Kamionkowski and K. Griest, Phys. Rept. 
{\bf 267} (1996) 195. 

\bibitem{ostriker1} Z. Haiman, R. Barkana and J.P.  Ostriker,
{\it Warm Dark Matter, Small Scale Crisis, and the High Redshift 
Universe}, astro-ph/0103050.

\bibitem{szalay} J.R. Bond and A.S. Szalay, Astrophys. J. {\bf
274} (1986) 443.

\bibitem{ostriker2}P. Bode, J.P. Ostriker and N. Turok, Astrophys.
J. {\bf 556} (2001) 93.

\bibitem{ostriker3} R. Barkana, Z. Haiman and J. P. Ostriker,
astro-ph/0102304.

\bibitem{first} 
S. Dodelson and L.M. Widrow, Phys. Rev. Lett. {\bf 72} (1994) 17.

\bibitem{dodelson} S. Colombi, S. Dodelson and L.M. Widrow,
Astrophys. J. {\bf 458} (1996) 1.

\bibitem{muller} 
C.M. M\"uller, Phys. Rev. {\bf D71} (2005) 047302.

\bibitem{DM7} 
M. Viel, J. Lesgourgues, M.G. Haehnelt, S. Matarrese, 
A. Riotto, Phys. Rev. Lett. 97 (2006) 071301; astro-ph/0605706.
% Can sterile neutrinos be ruled out as warm dark matter candidates?

\bibitem{Shap}  
M. Shaposhnikov, I. Tkachev, Phys. Lett. B639 (2006) 414;
%% The nuMSM, inflation, and dark matter.
%% e-Print: hep-ph/0604236

A. Boyarsky, A. Neronov, O. Ruchayskiy, M. Shaposhnikov, 
I. Tkachev, Phys. Rev. Lett. 97 (2006) 261302.
%% Where to find a dark matter sterile neutrino?
%% e-Print: astro-ph/0603660 

\bibitem{DM8}
K.A. Olive and J. Silk, Phys. Rev. Lett. 55 (1985) 2362; 

J. Ellis, D.V. Nanopoulos and S. Sarkar, Nucl. Phys. B 259 (1985) 175; 

J. Ellis, J.E. Kim and D.V. Nanopoulos, Phys. Lett. B 145 (1984) 181.

H. Pagels and J.R. Primack, Phys. Rev. Lett. 48 (1982) 223; 

S. Weinberg, Phys. Rev. Lett. 48 (1982) 1303.

D.V. Nanopoulos, K.A. Olive and M. Srednicki, Phys. Lett. B 127 (1983) 30; 

M.Yu. Khlopov and A.D. Linde, Phys. Let. B 138 (1985) 265; 

J. Ellis, E. Kim, and D.V. Nanopoulos, Phys. Lett. B 145 (1984) 181; 

R. Juskiewicz, J. Silk and A. Stebbins, Phys. Lett. B 158 (1983) 463; 

See, e.g., \cite{susyDM} for further references. 

\bibitem{DM6} 
N. Yoshida, A. Sokasian, L. Hernquist, V. Springel,
Astrophys. J. 591 (2003) L1-L4; astro-ph/0303622.
% Early structure formation and reionization in a warm 
% dark matter cosmology.

\bibitem{FlaFlu}
G. de Berredo-Peixoto, I. L. Shapiro and F. Sobreira,
Mod. Phys. Lett. {\bf 20A} (2005) 2723.

\bibitem{Juttner} F. J$\ddot{\rm u}$ttner, Ann. der Phys.
{\bf Bd 116} (1911) S. 145.

\bibitem{Kremer} C. Cercignani, G.M. Kremer,
{\sl The Relativistic Boltzmann Equation: Theory and Applications},
(Birkh\"{a}user, Basel, 2002)

\bibitem{FWC} R. Colistete Jr., J.C. Fabris, J. Tossa, W. Zimdahl,
Phys. Rev. {\bf D76} (2007) 103516;  arXiv:0706.4086[astro-ph]. 
%% Bulk Viscous Cosmology.

\bibitem{ma} C.-P. Ma and E. Bertschinger, 
Astrophysical J. {\bf 455}(1995) 7.

\bibitem{cole} S. Cole et al,
%\textit{The 2dF Galaxy Redshift Survey:
%Power-spectrum analysis of the final dataset and cosmological
%implications},
\textit{Mon. Not. Roy. Astron. Soc.} {\bf 362} (2005) 505;
\texttt{astro-ph/0501174}.

\bibitem{Analit-IFT}
 R. Aldrovandi, R. R. Cuzinatto and L. G. Medeiros, 
Found. Phys. {\bf 36} (2006) 1736, %% -1752 
gr-qc/0508073. 

\bibitem{Bardeen}
J.M. Bardeen, J.R. Bond, N. Kaiser and A.S. Szalay,
\textit{Astrophys. J}. {\bf 304} (1986) 15.
% THE STATISTICS OF PEAKS OF GAUSSIAN RANDOM FIELDS.

\bibitem{Martin} J. Martin, A. Riazuelo and M. Sakellariadou,
\textit{Phys. Rev.} {\bf D61} (2000) 083518.
 
\bibitem{CCwave} J.C. Fabris, I.L. Shapiro, J. Sol\`a,
%% {\it Density Perturbations for Running Cosmological Constant,}
JCAP {\bf 0702} (2007) 016, gr-qc/0609017.

\bibitem{Drukier} A.K. Drukier, K. Freese, D.N. Spergel, 
%% Detecting Cold Dark Matter Candidates.
Phys. Rev. {\bf D33} (1986) 3495. 

\bibitem{percival} W.J. Percival et al., 
Mon. Not. R. Astron.Soc. {\bf 327}, 1297(2001).

\bibitem{szapudi} M.C. Neyrinck and I. Szapudi, 
%% Mark C. Neyrinck and Istv\'an Szapudi, 
Mon. Not. R. Astron. Soc. {\bf 384}, 1221(2008).

\bibitem{sikivie} P. Sikivie, I.I. Tkachev and Y. Wang, 
Phys. Rev. Lett. {\bf 75} (1995) 2911;
Phys. Rev. {\bf D56}, 1863(1997).

\bibitem{abazajian} 
K. Abazajian, G.M. Fuller and M. Patel, 
Phys. Rev. {\bf D64} (2001) 023501.

\bibitem{mellier} L.C. Parker, 
{\it The masses and shapes of dark matter halos from 
galaxy-galaxy lensing in the CFHTLS},  arXiv:0707.1698.

\bibitem{binney} J. Binney and S. Tremaine, 
{\it Galactic dynamics} (Princeton University Press, 
Princeton, 2008).

\bibitem{walker} M.G. Walker et al, 
{\it Velocity dispersion profiles of seven dwarf 
spheroidal galaxies}, arXiv:0708.0010 [astro-ph].

\bibitem{dwarfs} 
G. Gentile, P. Salucci, U. Klein, G. L. Granato,
%% NGC 3741: dark halo profile from the most extended rotation curve
Mon. Not. Roy. Astron. Soc. {\bf 375} (2007) 199, %% -212 
astro-ph/0611355;

G. Gentile, A. Burkert, P. Salucci, U. Klein, F. Walter,
%% The dwarf galaxy DDO 47: testing cusps hiding in triaxial halos
Astrophys. Journ. {\bf 634} (2005) L145, %% -L148 
astro-ph/0510607. 

\bibitem{DMs1}     M. W. Goodman and E. Witten,
%% Detectability of certain dark-matter candidates
Phys. Rev. {\bf D31} (1985) 3059. %% - 3063

\bibitem{DMs2} 
A. K. Drukier, K. Freese and D. N. Spergel
 Phys. Rev. {\bf D33} (1986) 3495 - 3508
%% Detecting cold dark-matter candidates

\bibitem{DMs3}  M. Brhlik, L. Roszkowski,
%% WIMP Velocity Impact on Direct Dark Matter Searches
Phys. Lett. {\bf B464} (1999) 303, %% -310
hep-ph/9903468.

\bibitem{DMs4} 
P. Col\'{\i}n, A.A. Klypin and A.V. Kravtsov,
Velocity Bias in a Cold Dark Matter Model
Astroph. Journ. {\bf 539} (2000) 561. %% -569.

\bibitem{wilczek} K. Rajagopal, M.S. Turner and F. Wilczek, Nucl.
Phys. {\bf 358} (1991) 447.

\bibitem{DM9} 
J. Sommer-Larsen, A. Dolgov, Astrophys. J. 551 (2001) 608,
astro-ph/9912166.
% Formation of disk galaxies: warm dark matter and the 
% angular momentum problem.

\bibitem{grav1} Z.G. Berezhiani and  A.D. Dolgov, 
Phys. Lett. {\bf B375} (1996 ) 26.

\bibitem{grav2} N. Arkani-Hamed, S. Dimopoulos, G. Dvali 
and N. Kaloper, hep-ph/9911386. 

\bibitem{KM} 
M. Kaplinghat, Phys. Rev. {\bf D72} (2005) 063510.

\end{thebibliography}
\end{document}